%
\let\useblackboard=\iftrue
%
%
\newfam\black
\input harvmac.tex
\input tables.tex
%
\input epsf.tex
\ifx\epsfbox\UnDeFiNeD\message{(NO epsf.tex, FIGURES WILL BE
IGNORED)}
\def\figin#1{\vskip2in}
\else\message{(FIGURES WILL BE INCLUDED)}\def\figin#1{#1}\fi
\def\ifig#1#2#3{\xdef#1{fig.~\the\figno}
\midinsert{\centerline{\figin{#3}}%
\smallskip\centerline{\vbox{\baselineskip12pt
\advance\hsize by -1truein\noindent{\bf Fig.~\the\figno:} #2}}
\bigskip}\endinsert\global\advance\figno by1}
\noblackbox
\def\Title#1#2{\rightline{#1}
\ifx\answ\bigans\nopagenumbers\pageno0\vskip1in%
\baselineskip 15pt plus 1pt minus 1pt
\else
\def\listrefs{\footatend\vskip
1in\immediate\closeout\rfile\writestoppt
\baselineskip=14pt\centerline{{\bf
References}}\bigskip{\frenchspacing%
\parindent=20pt\escapechar=` \input
refs.tmp\vfill\eject}\nonfrenchspacing}
\pageno1\vskip.8in\fi \centerline{\titlefont #2}\vskip .5in}
 
scaled\magstep3
 
scaled\magstep3
 
scaled\magstep3
 
scaled\magstep3
 
scaled\magstep3
\ifx\answ\bigans\def\tcbreak#1{}\else\def\tcbreak#1{\cr&{#1}}\fi
\useblackboard
\message{If you do not have msbm (blackboard bold) fonts,}
\message{change the option at the top of the tex file.}

\font\blackboard=msbm10 scaled \magstep1
\font\blackboards=msbm7
\font\blackboardss=msbm5
\textfont\black=\blackboard
\scriptfont\black=\blackboards
\scriptscriptfont\black=\blackboardss

\else

\fi
%

%
\def\yboxit#1#2{\vbox{\hrule height #1 \hbox{\vrule width #1
\vbox{#2}\vrule width #1 }\hrule height #1 }}
\def\fillbox#1{\hbox to #1{\vbox to #1{\vfil}\hfil}}
\def\ybox{{\lower 1.3pt \yboxit{0.4pt}{\fillbox{8pt}}\hskip-0.2pt}}

\def\comments#1{}

\def\CN{{\cal N}}

\def\II{\relax{I\kern-.07em I}}

\def\IIB{{\II}B}

\def\inbar{\,\vrule height1.5ex width.4pt depth0pt}
\def\IZ{\relax\ifmmode\mathchoice
{\hbox{\cmss Z\kern-.4em Z}}{\hbox{\cmss Z\kern-.4em Z}}
{\lower.9pt\hbox{\cmsss Z\kern-.4em Z}}
{\lower1.2pt\hbox{\cmsss Z\kern-.4em Z}}\else{\cmss Z\kern-.4em
Z}\fi}
\def\IB{\relax{\rm I\kern-.18em B}}
\def\IC{{\relax\hbox{$\inbar\kern-.3em{\rm C}$}}}
\def\ID{\relax{\rm I\kern-.18em D}}
\def\IE{\relax{\rm I\kern-.18em E}}
\def\IF{\relax{\rm I\kern-.18em F}}
\def\IG{\relax\hbox{$\inbar\kern-.3em{\rm G}$}}
\def\IGa{\relax\hbox{${\rm I}\kern-.18em\Gamma$}}
\def\IH{\relax{\rm I\kern-.18em H}}
\def\IK{\relax{\rm I\kern-.18em K}}
\def\IP{\relax{\rm I\kern-.18em P}}
\def\pp{{\relax{=\kern-.42em |\kern+.2em}}}

\font\cmss=cmss10 \font\cmsss=cmss10 at 7pt
\def\IR{\relax{\rm I\kern-.18em R}}

\def\Nfour{{\CN\!=\!4}}

%
%
%
%

%
%

\Title{ \vbox{\baselineskip12pt\hbox{hep-th/9901143}
\hbox{TUW-99-01}
}}
{\vbox{
\centerline{String Corrections to the Hawking-Page Phase Transition}
}}

\centerline{Karl Landsteiner}\footnote{}{\tt
Research supported by the FWF under project P13125-PHY.}
\medskip
\centerline{Institut f\"ur theoretische Physik}
\centerline{Technische Universit\"at Wien, TU-Wien}
\centerline{Wiedner Hauptstra{\ss}e 8-10}
\centerline{A-1040 Wien, Austria}
\centerline{\tt landstei@brane.itp.tuwien.ac.at}
\bigskip
\bigskip

We compute the $O(\alpha^{\prime 3})$ corrections to the $AdS_5$ Black
hole metric in type \IIB\ string theory. Contrary to previous work in 
this direction we keep the Black Hole radius finite. Thus the topology
of the boundary is $S^3\times S^1$. We find the corrections
to the free energy and the critical temperature of the phase transition.

\Date{January, 1999}

\lref\juan{J. Maldacena, Adv. Theor. Math. Phys. 2 (1998) 231.}
\lref\wikleb{S. S. Gubser, I. R. Klebanov and A. M. Polyakov, Phys.Lett. B428 (1998) 105;

E. Witten, Adv.Theor.Math.Phys. 2 (1998) 253.}
\lref\wit{E. Witten, Adv.Theor.Math.Phys. 2 (1998) 505.}
\lref\gkp{S. S. Gubser, I. R. Klebanov and A. W. Peet, Phys.Rev. D54 (1996) 3915.}
\lref\gasi{G. Horowitz and S. Ross,J.High Energy Phys. 9804 (1998) 015.} 
\lref\kleb{I. R. Klebanov, ``From Threebranes to Large N Gauge Theories'', hep-th/9901018.}
\lref\gkt{S. S. Gubser, I. R. Klebanov and A. A. Tseytlin, Nucl.Phys. B534 (1998) 202.}
\lref\thpa{J. Pawelczyk and S. Theisen, J.High Energy Phys. 9809 (1998) 010.}
\lref\hapa{S. W. Hawking and D. N. Page, Commun. Math. Phys. 87 (1983) 577.}
\lref\gaoli{Y. Gao and M. Li, ``Large N Strong/Weak Coupling Phase Transition 
and the Correspondence Principle'', hep-th/9810053.}
\lref\bk{T. Banks and M. B. Green, J.High Energy Phys. 9805 (1998) 002.}
\lref\fota{A. Fotopoulos and T. R. Taylor, ``Comment on two loop free energy in
N=4 supersymmetric Yang-Mills theory at finite temperature'', hep-th/9811224.}


The study of large $N$ gauge theories received in the last year 
new input from Maldacena`s conjecture that type \IIB\  superstring
theory on asymptotically AdS - backgrounds is dual to the strong
coupling limit of large $N$ gauge theories \juan. Originally the conjecture
states that type \IIB\ string theory on $AdS_5\times S^5$ is dual to 
$\Nfour$ supersymmetric gauge theory with gauge group $SU(N)$ in 
the limit of large $N$ and strong 't Hooft coupling 
$\lambda = g^2_{YM}N$.
The string loop expansion corresponds to the $1/N$ expansion of the
gauge theory and
the sigma-model $\alpha^\prime$ expansion to the expansion in the 't Hooft
coupling. Denoting the scale of the $AdS_5\times S^5$ geometry by $\ell$, the 
AdS/CFT dictionary says ${\ell^4 \over \alpha^{\prime 2} } = {\lambda}$ and
$g_{string} = 4 \pi g^2_{YM}$. The gauge theory is thought to live on the 
conformal boundary of the $AdS$ space. Further specification of the
conjecture has been achieved in \wikleb, in particular it was shown how to 
calculate correlation functions in the boundary conformal field theory.
The conjecture was very soon generalized by Witten
to gauge theories at finite temperature \wit. He observed that replacing the
AdS space by the AdS Schwarzschild metric should correspond to have the
gauge theory at finite temperature on $S^3$. Taking the scaling limit of 
a large black hole one can make the radius of the three sphere infinite.
The metric obtained in this way coincides with the near horizon limit
of the near extremal threebrane solution \gasi. The relation between the
near extremal threebrane solution and gauge theories at finite temperature
has been studied already earlier by Klebanov et al.\gkp. In particular they
found that the Bekenstein-Hawking entropy of the threebrane solution 
reproduces the entropy of the $\Nfour$ gauge theory up to a relative factor
of $3\over 4$. It is now understood that this difference stems from the
fact, that the calculation in the gauge theory is done at vanishing coupling,
whereas the string calculation is relevant to the regime of infinite
coupling (see the recent review by Klebanov \kleb ). 
In \gkt\ the authors 
calculated the $O(\alpha^{\prime 3})$ correction to 
the entropy stemming from the $R^4$ term in the effective action of
type \IIB\ strings\foot{Corrections stemming from these terms have
first been conisered in context of the Ads/CFT correspondence in \bk.}. 
This calculation has actually a somewhat involved
history. It was first done using only the five dimensional limit of
the AdS Schwarzschild metric which is asymptotic to $S^1 \times R^3$
in two different ways. First the authors plugged in the uncorrected
expression for the metric in the $R^4$ term. In a second step they
also calculated the corrections to the metric. Quite surprisingly
the results for the free energy turned out to agree.
It was then pointed
out in \thpa\ that one actually has to consider the whole ten dimensional
metric including the $S^5$ factor. However, the thermodynamic quantities
turned out to be unchanged in comparison to the purely five dimensional
calculation. For the latter fact an explanation has been offered in an
appendix to \gkt. Using the AdS/CFT dictionary this correction translates
into the next to leading order term in a strong coupling expansion in 
$1 \over \lambda$ at infinite $N$. On general grounds one expects for
the entropy $S={2 \pi^2\over 3} N^2 V_3 T^3 f(\lambda)$. The strong coupling
expansion corresponding to the string theory regime gives then $f(\lambda) = {3 \over 4} + {45\over 32} \zeta(3) (2 \lambda)^{-3/2} + \cdots$. A recent field theory
two loop calculation gives the expansion around $\lambda=0$ as $f(\lambda) = 
1 - { 3 \over 2 \pi^2 } \lambda + \cdots$ \fota.

Let us turn now to the AdS Schwarzschild black hole which asymptotically
approaches $S^3 \times S^1$. The ten dimensional metric used as \IIB\  background
is given by 
\eqn\metric{ ds^2 = {r^2\over \ell^2} e^{-{10\over 3} C(r)} \left( e^{2 A(r) + 8 B(r)} d\tau^2 + e^{2 B(r)} dr^2 +
d \Omega_3^2 \right) + e^{2 C(r)} d\Omega_5^2 .}
We have chosen here the analog of the parameterization as in the appendix
of \gkt, which turns out to be very convenient for calculating the string
corrections. The $O(\alpha^{\prime\, 0})$ expression for the functions $A(r)$, 
$B(r)$ and $C(r)$ are
\eqn\abccassical{
\eqalign{
A(r) &=  -2 \log({r\over \ell}) + {5\over 2} \log({r^2\over \ell^2}+{r^4
\over \ell^4}-{r_0^2\over \ell^4});\cr
B(r) &=  -{1\over 2} \log({r^2\over \ell^2}+{r^4\over 
\ell^4}-{r_0^2\over \ell^4});\cr
C(r) &= 0.}}
Here $r_0$ is a parameter determining the mass of the black hole.
The thermodynamical properties of the four dimensional AdS Schwarzschild black 
hole have been investigated some years ago by Hawking and Page \hapa. For the five dimensional
case at hand things are completely parallel. We summarize the relevant facts.
The metric does not have a conical singularity if $\tau$ is compactified on a 
circle of radius $\beta = {2 \pi r_+ \ell^2 \over 2 r_+^2 +\ell^2}$. This 
defines the Hawking temperature $T={1 \over \beta}$ and $r_+$ is the largest
root of $r^2 \ell^2 + r^4 + r_0^4=0$. For a given temperature there are two solutions 
\eqn\bhradius{r_+ = {\pi T \ell^2\over 2}\left( 1 \pm \sqrt{1-{2 \ell^2\over \pi^2 T^2}}\;\right);}
the smaller
black hole is however unstable against decay into the larger one. The black hole solution exists only down to a minimal temperature 
$T_{min} = {\sqrt{2} \over \ell \pi}$. The action is
calculated by subtracting the value with $r_0=0$ which corresponds to empty
Anti-de-Sitter space. It changes sign at 
$r_+ = \ell$ ($T_c = {3\over 2\ell \pi}$) 
thereby signalling 
a phase transition. For $r_+ \le \ell$ the black hole is unstable against 
tunneling into
pure AdS space. Witten argued that this behavior corresponds to a large N
high/low temperature phase transition in the gauge theory on the boundary \wit.
At low temperature the theory is 
described by AdS-space. The free energy is of order one indicating a 
confining phase. At high temperature the AdS black hole geometry is thermodynamically favored. The free energy is of order $N^2$ indicating a deconfining
phase of the gauge theory. This high/low temperature phase transition occurs
only when the gauge theory lives on $S^3$. Actually, because of the conformal
symmetry of the $\Nfour$ gauge theory the behavior depends only on the
dimensionless quantity $\beta \over \ell$, measuring the ratio of the radius
of the circle to the radius of the three sphere. We are interested in
calculating the $O(\alpha^{\prime 3})$ corrections to the relevant 
thermodynamical quantities. This question has been addressed before in
\gaoli. There the authors did not consider corrections to the geometry
but just plugged in the uncorrected metric into the $R^4$ term.
This is a somewhat oversimplified procedure since it ignores the 
corrections to the geometry. It seems therefore worth to reconsider 
the problem taking into account also the $O(\alpha^{\prime 3})$ corrections
to the metric.  
For notational convenience we will set the AdS scale $\ell=1$, it can
be reintroduced easily by dimensional analysis. 

We use the action
\eqn\tenaction{ S = {1\over 16 \pi G_{10}} \int d^{10}x\, \sqrt{-g_{10}} \left(
R - {1\over 2} (\partial \Phi)^2 - {1\over 4. 5!} (F_5)^2 + 
\gamma e^{-{3\over 2}\Phi} W \right).}
with 
\eqn\defz{\gamma = {1\over 8} \zeta(3) \alpha^{\prime 3}.}
$W$ denotes the $R^4$ term and can be expressed as a particular contraction
of four Weyltensors
\eqn\defW{ W = C_{abcd}\,C^{ebcf}\,C^{a}\,_{ghe}\,C_{f}\,^{ghd} + {1\over 2} 
C_{adbc}\,C^{efbc}\,C^{a}\,_{ghe}\,C_{f}\,^{ghd}\,.}
For the metric at $O(0)$ we have $W = {180 r_0^8 \over r^{16}}$.
The action for \IIB\ five form field should be understood in the sense that
one first varies the action and then implements the self-duality condition at
the level of the equation of motion. Such a procedure is convenient for
purposes of dimensional reduction. In particular using the ansatz \metric\ we
obtain the one dimensional action
\eqn\oneaction{ 
\eqalign{S =& {Vol(S^3)Vol(S^5) \beta \over 16 \pi G_{10}}
\int dr\, \left[ r^5 e^{A+5B-{16\over 3} C}(20-8e^{-8C}) +
{10\over 3}(r^3  C^\prime e^{A+3B})^\prime + \right.\cr
& e^{A+3B}\Bigl(6r^3 e^{2B}-4r-8r^2A^\prime-2r^3(A^\prime)^2 - 24 r^2 B^\prime - 
14 r^3 A^\prime B^\prime - 24 r^3(B^\prime)^2 - \cr
&\left. {40\over 3} (C^\prime)^2 - 2 r^3 A^{\prime\prime} - 8 r^3 B^{\prime\prime} \Bigr) + \gamma \hbox{W-contribution} 
\right].}}
As already in the $R^3$ case the W-contribution is to long to be displayed 
here (approx. 1.5 MB Mathematica output).
Let us concentrate for a moment on the scale factor for the $S^5$. Since it 
will be of order $\gamma$ we find that its contribution to the action
at this order is given by the total derivative term solely.
\eqn\contribC{ S_{C} = e^{A+3B} r^3 C^\prime \Bigr|_{r_+}^{\infty}.}
This expression vanishes at the horizon $r_+$. At infinity it vanishes also
provided
$C(r)$ goes to zero at least as fast as $1\over r^5$.
We set now $A(r) = A_0(r) + \gamma A_1(r)$, etc. The equations of motion
following from the action are
\eqn\eoms{\eqalign{
\left[(r^4+r^2-r_0^2)B_1\right]^\prime + {10 r_0^6\over r^{13}} (171r_0^2 - 160 r^2 - 144 r^4) =0\,;\cr
(r^4+r^2-r_0^2)A_1^\prime - 10 (2r^3+r) B_1 + {10 r_0^6\over r^{13}} (576 r^4 + 565 r^2 - 711 r_0^2) = 0\,;\cr
\left[(r^5+r^3-r r_0^2)C_1^\prime\right]^\prime - 32 r^3 C_1 + {135 r_0^8\over 2 r^{13}} =0\,;\cr
\left[(r^4+r^2-r_0^2)\Phi_1^\prime \right]^\prime - {270 r_0^8 \over r^{13}} =0\,.
}}
Despite much effort we did not find a closed solution for $C_1$. However,
as we just saw, this field does not contribute to the action if it vanishes
fast enough at infinity. For very large $r$ the differential equation for
$C_1$ becomes
\eqn\Cinf{ (r^5 C_1^\prime)^\prime - 32 r^3 C_1 =0\,.}
From this we find  $C_1(r) \approx {c_1 \over r^8} + c_2 r^4$. As boundary
condition we demand that the metric should still be asymptotically AdS and
thus $c_2=0$. This shows that $C_1(r)$ vanishes indeed fast enough and does
not contribute to the action. The solutions for the other equations are
\eqn\sols{\eqalign{
A_1(r) =& (r^4+r^2-r_0^2)^{-1}\left({1360 r^2 r_0^6 + 1560 r^4 r_0^6 - 1185 r_0^8\over 
2r^{12} } - {25 r_0^6 (15r_0^2- 8 r_+^2)\over 2 r_+^{12}}\right)\,;\cr
B_1(r) =& (r^4+r^2-r_0^2)^{-1}\left( {285 r_0^8 - 360 r^4 r_0^6 - 320 r^2 r_0^6 \over 2 r^{12}} + {5r_0^6(15r_0^2-8r_+^2)\over 2r_+^{12}}\right)\,; \cr
\Phi_1(r) =& {45\over 4 r_0^4 (r_+^2+1)} (3r_0^4+4r_0^2+1)\log\left(
{2(r^2+r_+^2+1)\over r^2}\right) - 
{3\over 16 r_0^4 r^{12}} \bigl[ (180 r_0^4 + 240 r_0^2 + 60)r^{10} +\cr  
& (30r_0^6+90r_0^4+30r_0^2)r^8 + (40r_0^6+20r_0^4)r^6 + 15r_0^6(r_0^2+1)r^4 + 12r_0^8r^2 + 10 r_0^{10}\bigr]\;.}} 
Demanding that the corrected metric does not have a conical singularity
at the horizon $r_+$ we find the period $\beta$ and thus the temperature
$T={1\over \beta}$ to be 
\eqn\temp{ T = {2 r_+^2 + 1\over 2\pi r_+} \bigl( 1 + \gamma {10 (r_+^3+1)^3
(3r_+^2-1) \over r_+^6 (2r_+^2 + 1)} \bigr).}
The action at $O(\gamma)$ is
\eqn\action{ S_R = {\beta Vol(S^3)Vol(S^5) \over 16 \pi G_{10}} \bigl( -2r^4 - \gamma30 { r_0^6 (4r^2+r_0^2) \over r^{12} } \bigr) \Bigl|_{r_+}^{r_{max}} \,.}
We regularize this expression in the usual way by subtracting the action
for the metric with $R=0$ and the same asymptotic geometry. This condition
demands that the period of the empty AdS space $\beta_0$ for large 
radii is related to
the period of the AdS black hole $\beta$ through
\eqn\betao{\beta_0 \approx \beta \bigl( 1-{r_0^2\over 2 r^4} - \gamma {5 (r_+^2+1)^3(7+15 r_+^2)\over 2 r_+^4} {1\over r^4} + O({1\over r^5}) \bigr)\,.}
We end up with the following expression for the free energy $F = {1\over \beta}
\hbox{lim}_{r_{max}\rightarrow \infty}(S_R - S_{R=0}) $
\eqn\regaction{F =
{Vol(S^3) Vol(S^5) \over 16\pi G_{10}} \Bigl[ r_+^4 - r_+^2 + \gamma
{(r_+^2+1)^3 (75r_+^2 -5) \over r_+^4} \Bigr]\,.} 
The correction to the critical black hole radius that follows from this
expression is
\eqn\rpluscrit{ r_+^c = 1 - 280 \gamma\,;}
and the corrected critical temperature is
\eqn\tempcrit{ T_c = {3\over 2\pi} - \gamma {60\over \pi}\,.}
We finally want to express the free energy as function of the temperature.
To do this we need to invert \temp. We write $r_+ = f_0(T) + \gamma f_1(T)$
where $f_0(T)$ is given by the larger value in \bhradius\ and $f_1(T)$ turns
out as
\eqn\fone{ f_1(T) = - { 20 \left( 1 + T^2\pi^2 \left( 1 + \sqrt{1-{2\over T^2\pi^2}}\right)\right)^3 \left(-5 + 3\pi^2 T^2 \left(1 + \sqrt{1-{2\over T^2\pi^2}}\right)\right) \over \pi^5 T^5 \left(1 + \sqrt{1-{2\over T^2\pi^2}}\right)^5 \left( -2 + \pi^2 T^2 \left(1 + \sqrt{1-{2\over T^2\pi^2}}\right)\right)\,.}}
Plugging this into \regaction\ we obtain the free energy as function of the
temperature
\eqn\freeenergy{ \eqalign{
F =& -{Vol(S^3) Vol(S^5) \over 16\pi G_{10}} \Bigl[
{1\over 8}\pi^2 T^2 \left(1+\sqrt{1-{2\over \pi^2 T^2}}\right)^2 \left(
-3 + \pi^2 T^2 \left(1+\sqrt{1-{2\over \pi^2 T^2}}\right)
\right)\cr +
& \gamma {15\over 4} \left( 1 + 2\pi^2 T^2 \left(-14 + 3 \sqrt{1-{2\over T^2\pi^2}}\right)
+ \pi^4 T^4 \left( 34 -30 \sqrt{1-{2\over T^2\pi^2}} \right) \right)\Bigr]\,.
}}
The entropy is now given by $S = - {\partial F \over \partial T}$ and
coincides exactly with the expression given in \gaoli. It is very interesting
that a careful calculation taking into account also the corrections to
the metric gives the same results as just plugging in the uncorrected metric
into the $R^4$ term. Noting that the corrections are rather more complicated
that in the flat case considered in \gkt\ and \thpa\ this seems even more a 
miracle. So the features
of the geometry at $O(\alpha^{\prime 3})$ are not relevant to its 
thermodynamics at this order. It would be interesting to have a better
physical understanding of this property. Another interesting point would
be to compare these expressions with a two loop calculation in field theory.


\bigskip
{\bf Acknowledgements}

I would like to thank H. Balasin and E. Lopez for discussions.

\listrefs
\end